\documentclass[acmsmall,screen]{acmart}

\makeatother
\AtBeginDocument{%
  }

\setcopyright{acmlicensed}

\acmDOI{XXXXXXX.XXXXXXX}

\usepackage{siunitx}
\usepackage{dashrule}

\usepackage[ruled,vlined]{algorithm2e}
\usepackage{float}
\usepackage{amsthm}
\usepackage{tabularx}
\usepackage{graphicx}
\usepackage{svg}
\usepackage{romannum}

\newtheorem{definition}{Definition}
\usepackage{tcolorbox}
\usepackage{minted}  
\usepackage{xcolor} 
\usepackage{listings}
\usepackage{multicol}
\usepackage{pgfplots}
\pgfplotsset{compat=1.18}
\usepackage{tikz}
\usepackage{ragged2e}
\usepackage{makecell}
\usepackage{booktabs, multirow,diagbox}
\tcbset{
  mybox/.style={
    colframe=gray!70,
    colback=white,
    colbacktitle=gray!15,  
    coltitle=black,        
    fonttitle=\bfseries\small,
    boxrule=0.5pt,
    arc=2pt,
    left=3pt,right=3pt,top=3pt,bottom=3pt,
    before skip=3pt, after skip=3pt,
    title filled          
  }
}
\definecolor{bottlegreen}{RGB}{0,20,252}
\usepackage{caption}
\lstdefinestyle{ath}{
  basicstyle=\ttfamily\scriptsize\color{bottlegreen},
  breaklines=true,          
  keepspaces=true,
  showstringspaces=false,
  columns=fullflexible,
  xleftmargin=0pt,
  aboveskip=2pt
}

\newtcolorbox{rqbox}{               
  colback=gray!10,          
  colframe=black,           
  boxrule=0.6pt,           
  arc=3mm,                 
  left=8pt,right=8pt,top=6pt,bottom=6pt, 
}

\begin{document}

\title{Analyzing and Mitigating Surface Bias in Code Evaluation Metrics}

\author{Simantika Bhattacharjee Dristi}

\affiliation{%
  \institution{University of Virginia}
  \city {Charlottesville}
  \country{USA}}
\email{nwc8gr@virginia.edu}
\orcid{0009-0007-5511-9307}

\author{Matthew B. Dwyer}
\affiliation{%
  \institution{University of Virginia}
  \city {Charlottesville}
  \country{USA}}
\email{matthewbdwyer@virginia.edu}
\orcid{0000-0002-1937-1544}

\begin{abstract}
With the increasing popularity of large language models (LLMs) and LLM-based agents, reliable and effective code evaluation metrics (CEMs) have become crucial for progress across several software engineering tasks. While popular benchmarks often provide test cases to assess the correctness of generated code, crafting and executing test cases is expensive. Reference-based CEMs provide a cheaper alternative by scoring a candidate program based on its \textit{functional similarity} to a reference. Although prior research has focused on reporting the weak correlation between these CEMs and functional correctness, the causes are only assumed, and plausible solutions remain unexplored. In this work, we critically evaluate four state-of-the-art reference-based CEMs, revealing their strong bias towards surface-level features rather than code functionality. Despite this \textit{surface bias}, current evaluation datasets for these CEMs rarely include code pairs that are surface-similar yet functionally dissimilar, or functionally similar yet surface-dissimilar.
\par To mitigate this gap, we propose \textbf{LoCaL (\underline{Lo}oks \underline{Ca}n \underline{L}ie)}, a CEM evaluation benchmark, with \textbf{3117} code pairs at both the method and program levels. Each pair is labeled with a functional similarity score and aims to target regions where CEMs are likely to perform poorly. The functional similarity scores are calculated through differential fuzzing, which eliminates the need for predefined test cases and, at the same time, improves the reliability of the scores by executing an order of magnitude more tests than prior work. We find that all four CEMs show significant performance degradation on LoCaL, compared to the baselines. Finally,  based on our findings, we draw the implication that exposing CEMs to LoCaL-like data might facilitate the development of metrics that are robust to \textit{surface bias}. 
\end{abstract}



\maketitle
\pagenumbering{arabic}

\section{Introduction}

 Automated code generation has emerged as a focal point in agent-driven software engineering (SE), primarily due to its ability to significantly reduce development cost and enhance programming efficiency \cite{inproceedings2, unknown}. This, in turn, has led to rapid advancements in code generation techniques \cite{inbook1, inproceedings, doi:10.1126/science.abq1158, 10.1145/3660810, 10795375}, with tools like ChatGPT \cite{chatgpt} and Github Copilot \cite{chen2021evaluatinglargelanguagemodels} being widely adopted in practice. 
Code-generation approaches have been developed using \textit{code evaluation metrics} (CEM) that assess the quality of machine-generated code. Such metrics, though developed for code generation, are equally essential for a broad range of software engineering tasks, including code completion, code translation, code refactoring, code optimization, and automated program repair \cite{Zhu_Suresh_Reddy_2022, pan2024codevbenchllmsunderstanddevelopercentric, 10704582, 11024270, 10.1145/3691620.3695537}. 
While prior research in code evaluation has enabled initial progress across a range of agent-driven SE tasks, improvement in CEMs has the potential to drive further advances in these areas.
\par The evaluation of generated code has long relied on test cases to check functional correctness \cite{fi16060188, chen2022codetcodegenerationgenerated, 10.5555/3666122.3667065}. Reference-based static metrics \cite{ren2020codebleumethodautomaticevaluation, 10.1145/3551349.3556903, zhou2023codebertscore, 10.1145/3695991} are a more recent approach that estimates the functional similarity between a reference and a generated code without execution. These metrics are attractive because they are test-case agnostic and offer faster analysis than execution-based methods \cite{10.1145/3747588, ren2020codebleumethodautomaticevaluation, 10.1145/3551349.3556903, zhou2023codebertscore, 10.1145/3695991},
but they lack the ground truth of test-based equivalence approaches.  To date, there has been little research focused on critically evaluating the reliability of these reference-based static code evaluation metrics.
\par In many SE tasks, the key criteria for assessing code is related to its functional behavior, and to support such tasks, CEMs must capture code semantics rather than surface-level textual or structural similarity \cite{EVTIKHIEV2023111741}. In real-world scenarios, code that appears nearly identical can behave differently, while code that looks different may be functionally identical.  
For example, as shown in Fig. \ref{fig:example_opts_muts}, code (a) and its mutated variant (b) appear almost alike but diverge drastically in their execution results, while (a) and its optimized variant (c) look substantially different but preserve the functional behavior. These cases highlight a fundamental question that must be asked while evaluating reference-based static CEMs -- \textbf{How reliably do these CEMs capture functional similarity in scenarios where surface-level resemblance and functional behavior diverge significantly?}
\begin{figure*}[t]
\centering
\setlength{\fboxrule}{1pt}
\begin{minipage}[t]{0.32\textwidth}
\begin{tcolorbox}[mybox, title={\centering (a)}, colback=white, height=3.3cm]
\begin{lstlisting}[style=ath, escapeinside=@@]
@\textcolor{orange}{def}@ is_palindrome(text: str) -> bool:
    @\textcolor{orange}{for}@ i in range(len(text)):
        @\textcolor{orange}{if}@ text[i] @\fcolorbox{red}{white}{!=}@ text[len(text) - 1 - i]:
            @\textcolor{orange}{return}@ False
    @\textcolor{orange}{return}@ True
\end{lstlisting}
\end{tcolorbox}
\end{minipage}
\hfill
\begin{minipage}[t]{0.32\textwidth}
\begin{tcolorbox}[mybox, title={\centering (b)}, colback=white, height=3.3cm]
\begin{lstlisting}[style=ath, escapeinside=@@]
@\textcolor{orange}{def}@ is_palindrome(text: str) -> bool:
    @\textcolor{orange}{for}@ i in range(len(text)):
        @\textcolor{orange}{if}@ text[i] @\fcolorbox{red}{white}{==}@ text[len(text) - 1 - i]:
            @\textcolor{orange}{return}@ False
    @\textcolor{orange}{return}@ True
\end{lstlisting}
\end{tcolorbox}
\end{minipage}
\hfill
\begin{minipage}[t]{0.32\textwidth}
\begin{tcolorbox}[mybox, title={\centering (c)}, colback=white, height=3.3cm]
\begin{lstlisting}[style=ath, escapeinside=@@]
@\textcolor{orange}{def}@ is_palindrome(input_str: str) -> bool:
    @\textcolor{orange}{return}@ input_str == input_str[::-1]
\end{lstlisting}
\end{tcolorbox}
\end{minipage}

\vspace{0.6em}

\centering
\scriptsize
(a, b) : Surface Similarity = 0.99; Functional Similarity = 0.00\hfill
(a, c) : Surface Similarity = 0.48; Functional Similarity = 1.00
\vspace{-2mm}
\caption{Examples of surface–semantic divergence in code. Red borders mark how trivial changes in the token-level can flip code semantics entirely.}
\label{fig:example_opts_muts}
\end{figure*}

\par In this paper, we evaluate four popular reference-based code evaluation metrics: CodeBLEU \cite{ren2020codebleumethodautomaticevaluation}; CrystalBLEU \cite{10.1145/3551349.3556903}; CodeBERTScore \cite{zhou2023codebertscore}; and CodeScore \cite{10.1145/3695991}, on functionally equivalent code pairs that have varying degrees of surface similarity. In Section~\ref{RQ1}, we report experimental results demonstrating that all four metrics have \textbf{moderate to very strong positive correlation} with surface similarity. P-values, as low as $10^{-324}$, provide statistically significant evidence that these metrics are biased towards how the code looks rather than what it does. We refer to this as the \textbf{\textit{surface bias}} of a code evaluation metric.
 
\par We further analyze \textit{distinguishability} (detailed in Section \ref{distinguishability}), a meta-metric that evaluates how well a code evaluation metric can distinguish functionally equivalent pairs from the non-equivalent ones. CrystalBLEU 
has been reported to outperform other baselines in terms of distinguishability \cite{10.1145/3551349.3556903}.
We provide empirical evidence in Section \ref{RQ2} that the distinguishability of CEMs considered substantially degrades in the presence of surface bias, with \textbf{CrystalBLEU itself showing the largest drop of 95.3\%}. We argue that the performance drop arises because no prior evaluation has explicitly evaluated CEMs in settings where surface similarity and functional behavior move in opposite directions. 
To address this need, in Section~\ref{sec:SFD-DFS} we define two contrasting evaluation regions: \textit{Similar Form, Different Semantics} (SFD), which includes code pairs where the surface similarity largely exceeds functional similarity, and \textit{Different Form, Similar Semantics} (DFS), which provides for code pairs where functional similarity dominates over surface similarity. As we show in Section \ref{RQ3}, existing work \cite{10.1145/3695991} fails to adequately cover the DFS and SFD regions, revealing a concerning gap in how CEMs have been assessed. 

\par
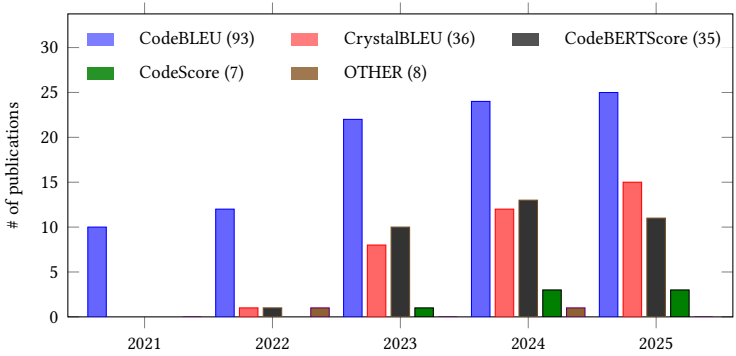
\begin{figure}[t]
\centering
\begin{tikzpicture}
\begin{axis}[
    ybar,
    bar width=7pt,
    ymin=0,
    font=\scriptsize,
    ylabel={\# of publications},
    symbolic x coords={2021,2022,2023,2024,2025},
    xtick=data,
    enlarge x limits=0.15,
    enlarge y limits={upper, value=0.35},
    width=0.75\linewidth,
    height=0.4\linewidth,
    ytick distance=5,  
    legend style={
    at={(0.02,0.98)}, anchor=north west,
    legend columns=3,
    font=\scriptsize,
    draw=none,
    fill=white, fill opacity=0.85, text opacity=1,
    inner sep=2pt, row sep=2pt, column sep=8pt,
    /tikz/every even column/.style={column sep=12pt}
  },
   legend image code/.code={%
    \draw[#1,draw=none,fill] (0cm,-0.08cm) rectangle (0.35cm,0.08cm);
  },
  legend cell align=left
]

\addplot+[fill=blue!60] coordinates
 {(2021,10) (2022,12) (2023,22) (2024,24) (2025,25)};
\addlegendentry{CodeBLEU (93)}

\addplot+[fill=red!60] coordinates
 {(2021,NaN) (2022,1) (2023,8) (2024,12) (2025,15)};
\addlegendentry{CrystalBLEU (36)}

\addplot+[fill=black!80] coordinates
 {(2021,NaN) (2022,1) (2023,10) (2024,13) (2025,11)};
\addlegendentry{CodeBERTScore (35)}

\addplot+[fill=green!50!black] coordinates
 {(2021,NaN) (2022,NaN) (2023,1) (2024,3) (2025,3)};
\addlegendentry{CodeScore (7)}

\addplot+[fill=brown!75!black] coordinates
 {(2021,0) (2022,1) (2023,0) (2024,1) (2025,0)};
\addlegendentry{OTHER (8)}

\end{axis}
\end{tikzpicture}

\vspace{-4mm}
\caption{Publications per year (2021–2025) in which each CEM is used as part of the evaluation. Legend parentheses show totals across all years. Missing bars indicate no usage or metrics not yet introduced.}
\label{fig:bar_chart}
\end{figure}

 We are not the first to observe flaws in reference-based static CEMs \cite{EVTIKHIEV2023111741, naik2024limitationsembeddingbasedmethods, 10685214}, yet they remain widely used for code evaluation in recent publications. 
We quantify the recent adoption of popular reference-based static CEMs through a systematic literature search. We query both Google Scholar and arXiv for each metric that we evaluate (CodeBLEU, CrystalBLEU, CodeBERTScore, and CodeScore), limiting the search between 2021 and 2025. We then manually filter the results to include only publications that employ the metric as part of the evaluation, rather than just mentioning it. If an included paper also used a different CEM, that CEM was counted under ``OTHER.”
Figure~\ref{fig:bar_chart} plots the number of papers using each metric for evaluation over the past five years.
As the figure illustrates, CodeBLEU is the most frequently used metric, with 93 prior works relying on it for evaluation. This is concerning because our findings in Section~\ref{sec:experiments} indicate CodeBLEU is also among the most sensitive to \textit{surface bias}.  This highlights the urgent need for more robust metrics and frameworks to evaluate them, but no benchmark currently exists with sufficiently challenging samples to assess code evaluation metrics effectively. 

\par To mitigate this gap, we introduce \textbf{LoCaL (\underline{Lo}oks \underline{Ca}n \underline{L}ie)}, a benchmark comprising \textbf{3117} Python code pairs at both method-level and program-level. Each pair in LoCaL is carefully curated to target either the SFD or DFS region and is annotated with a functional similarity score computed using a differential fuzzing-based pipeline. To the best of our knowledge, LoCaL is the first large-scale dataset specifically designed to quantify functional similarity between code pairs using execution results. LoCaL builds on widely adopted benchmarks: HumanEval \cite{chen2021evaluatinglargelanguagemodels} and MBPP \cite{austin2021programsynthesislargelanguage} for method-level pairs, and APPS \cite{hendrycks2021measuring} and PIE \cite{shypula2024learning} for program-level pairs. As described in Section~\ref{LoCal_consttruction}, functional similarity scores are computed through a structured three-step process. First, for a reference code in the existing benchmark, two types of variants are generated: optimized variants using GPT-4 \cite{openai2024gpt4technicalreport} guided code optimization and mutated variants via injecting minimal changes to the source code that can alter program behavior. Second, a large language model is used to infer constraints over the inputs, which are further manually verified to ensure that any inter-dependencies or relational conditions between the inputs/arguments are respected. Finally, a fuzzer generates a diverse set of test inputs within the defined input space, and each input is executed on both the reference and its paired variant to compute a similarity score based on the proportion of matching outputs. 
\par Prior work~\cite{10.1145/3695991} introduced
CodeScore, which computes functional similarity using a fixed test suite, whereas our approach uses differential fuzzing to generate test inputs that are diverse and that adhere to the problem specification. Moreover, when CodeScore used just over 100 test cases per pair, hand ``built'' as per the authors' description \cite{10.1145/3695991}, LoCaL automatically generated 1,000–2,000 test cases per pair and can be scaled to further boost coverage of function behavior and to support new benchmark pairs without human intervention.
LoCaL stands out as a benchmark for several key reasons:
(1) The differential fuzzing approach \textbf{eliminates the reliance on pre-defined test cases} by dynamically generating test cases during execution;
(2) Instead of generating inputs by random byte fuzzing, LoCaL \textbf{uses verified argument constraints} to generate valid and meaningful test inputs; and
(3) LoCaL \textbf{includes both method-level and program-level code pairs}, offering a broader evaluation of CEMs across different tasks. 

Experiments demonstrate that LoCal is highly effective in revealing the surface bias of code evaluation metrics, yielding \textbf{+0.47 ($\mathbf{+\approx158\%}$) mean absolute error (MAE) } on average, compared to the baseline.

Moreover, in an exploratory study, we find that injecting LoCaL samples into CodeScore's training set can reduce its error on held-out LoCaL data by up to $\approx 85\%$ while retaining comparable performance on its original inference dataset. Although exploratory, the gain suggests 
the potential for LoCaL-like data to counter the \textit{surface bias} that inhibits today's code evaluation metrics. 
\par The primary contributions of this paper can be summarized as follows:
\begin{itemize}
    \item We conduct the first in-depth statistical analysis of how \textit{surface bias} impairs the performance of widely used code evaluation metrics.
     \item We define two contrasting evaluation regions — SFD (Similar Form, Different Semantics)  and DFS (Different Form, Similar Semantics) to isolate and target the areas where \textit{surface bias} is most likely to mislead the code evaluation metrics. 
    \item We present LoCaL (\underline{Lo}oks \underline{Ca}n \underline{L}ie), the first large-scale benchmark of both method-level and program-level code pairs, annotated with functional similarity scores obtained via differential fuzzing.
    \item  We perform a comprehensive evaluation of four state-of-the-art CEMs on LoCaL, revealing concerning gaps in their ability to understand code semantics. 
    \item We provide LoCaL as an open and extensible resource to support the development and validation of future, semantics-aware evaluation metrics for code.
\end{itemize}

\section{Background and Related Work}
\subsection{Code Evaluation Metrics (CEMs)}
The primary goal of a code evaluation metric (CEM) is to assess the quality of a machine-generated code. While predominantly applied to evaluating functional correctness \cite{ren2020codebleumethodautomaticevaluation, 10.1145/3551349.3556903, zhou2023codebertscore, 10.1145/3695991, du2024mercurycodeefficiencybenchmark, zhuo-2024-ice, tong2024codejudgeevaluatingcodegeneration}, CEMs have also been utilized to measure efficiency of the generated code \cite{du2024mercurycodeefficiencybenchmark, peng2025coffecodeefficiencybenchmark}. Code evaluation metrics broadly fall into three categories: \textit{Execution-based metrics}, \textit{Reference-based static metrics}, and \textit{Instruction-based metrics}.

\textbf{Execution-based metrics}, such as pass@k \cite{chen2021evaluatinglargelanguagemodels} and n@k \cite{doi:10.1126/science.abq1158}, execute test cases on the generated code to assess its functional correctness. While these metrics provide stronger evidence of correctness, they are inapplicable to problems that do not come with pre-defined test cases. 

\textbf{Reference-based static metrics}, on the other hand, compare the generated code with a reference solution without executing the code or analyzing run-time behavior. These metrics can further be classified into two subtypes.
\textit{Match-based metrics}: Earlier approaches like BLEU \cite{10.3115/1073083.1073135}, METEOR \cite{banerjee-lavie-2005-meteor} treated code snippets as natural language and calculated a similarity score using n-gram match between the reference and the generated code. More recent match-based metrics such as CodeBLEU \cite{ren2020codebleumethodautomaticevaluation}, CrystalBLEU \cite{10.1145/3551349.3556903}, and RUBY \cite{Tran_2019} extend the earlier approaches by considering code-specific syntactic and semantic features when measuring similarity.
\textit{Model-based metrics}: CEMs like COMET \cite{rei-etal-2020-comet}, CodeBERTScore \cite{zhou2023codebertscore}, and CodeScore \cite{10.1145/3695991}, leverage pre-trained language models to comprehend and compare code semantics, going beyond mere lexical overlap.

\textbf{Instruction-based metrics}, including ICE-Score \cite{zhuo-2024-ice}, CODE-DITING \cite{yang2025codeditingreasoningbasedmetricfunctional}, and, CodeJudge \cite{tong2024codejudgeevaluatingcodegeneration}, have more recently leveraged structured instructions given to an \textit{LLM-judge} for code evaluation. Although these metrics are reference-free, their effectiveness is constrained by the core LLM's limitations in comprehending code semantics \cite{haroon2025accuratelylargelanguagemodels, Nacl}. Moreover, instruction-based CEMs require a natural language description of the problem, which may not always be available.\\
In this paper, we focus on the reference-based static CEMs only, with the central question being whether they can reliably measure the degree of \textit{functional similarity} between two different code implementations. Accordingly, we do not consider instruction-based metrics since they evaluate a code implementation against its natural language description.

\subsection{Differential Fuzzing} 
Differential testing minimizes the cost of evaluating test results by running the same input on two (or more) implementations of the same problem and flagging any behavioral mismatch as a potential bug \cite{mckeeman1998differential}. Differential fuzzing enhances this approach with an automated, large-scale input generator to expose functional mismatch across multiple implementations \cite{10.1145/3238147.3241537}. The input generator, i.e., the fuzzer, typically starts with a tiny seed corpus of valid test inputs, aggressively mutates them to generate new test inputs, and adds interesting inputs back to the corpus to be mutated again. The usefulness of differential fuzzing is two-fold: it eliminates the need for explicit oracles, and it does not require a pre-defined corpus of test inputs. While it has previously been applied to detecting vulnerabilities in deep learning libraries and compilers \cite{df1, df2, df3}, no prior work has leveraged differential fuzzing for code functional similarity analysis.

\subsection{Evaluating the Evaluators}
Research in evaluating CEMs has primarily focused on showing their poor correlation with human judgment or functional correctness \cite{Tran_2019, hendrycks2021measuring, 10.5555/3454287.3455353, EVTIKHIEV2023111741, naik2024limitationsembeddingbasedmethods}, but the underlying reasons behind this limitation remain underexplored. Despite its widespread use, several studies have shown BLEU \cite{10.3115/1073083.1073135} to be a weak reflector of functional correctness \cite{Tran_2019,hendrycks2021measuring, 10.5555/3454287.3455353}. Evtikhiev et al. \cite{EVTIKHIEV2023111741} were the first to systematically evaluate six CEMs - BLEU, ROUGE-L \cite{lin-2004-rouge}, METEOR \cite{banerjee-lavie-2005-meteor}, ChrF \cite{popovic-2015-chrf}, CodeBLEU \cite{ren2020codebleumethodautomaticevaluation}, and RUBY \cite{Tran_2019}, revealing their poor alignment with human judgment. However, since the study lacks execution-based validation, it fails to provide definitive evidence of the metrics' inadequacy. 
\par Another line of work, though not centered around CEMs, highlights the weakness of LLMs in understanding code semantics \cite{haroon2025accuratelylargelanguagemodels, Nacl}. For example, Maveli et al. \cite{Nacl} used code clones, a type of code refactoring, to introduce  Semantic Preserving Mutations (SPM) in source code and assess LLMs' ability to capture 0/1 equivalence. However, later studies suggest code evaluation should go beyond binary (0/1) decision, since partially correct code can aid further debugging \cite{10.1145/3491101.3519665, 10.1145/3586030}. Moreover, we found that applying automated code refactoring tools \cite{rewrite_python, pyrefact} generally results in code edits that were localized and minimal and were thus unable to 
expose surface bias.
\par Limited research has focused on developing benchmarks that particularly target execution-less CEMs. To our knowledge, the test split of CodeScore \cite{10.1145/3695991} is the only public benchmark offering a continuous functional similarity score between two codes in a pair. However, since the code pairs in this benchmark were not designed to expose any particular weakness of CEMs, they offer limited challenge. Additionally, with just over 100 test cases per pair and no checks for argument dependency in the test inputs, the reliability of its scores remains uncertain.

\section{Approach}
We begin this section by providing an overview of the selected metrics, followed by our approach to quantifying the impact of \textit{surface bias} on those metrics. We then discuss the construction of our benchmark, LoCaL, and the algorithm used to define the SFD and DFS regions. 

\subsection{Selection of Metrics} We chose CodeBLEU, CrystalBLEU, CodeBERTScore, and CodeScore as our metrics of study to cover the two main types of reference-based static metrics:  match-based (CodeBLEU and CrystalBLEU) and model-based (CodeBERTScore and CodeScore). All of these metrics were specifically designed for code evaluation, unlike several others \cite{10.3115/1073083.1073135, lin-2004-rouge, banerjee-lavie-2005-meteor} that, while commonly applied to code, originate from natural language tasks. Moreover, as Fig. \ref{fig:bar_chart} shows, they have been widely used for code evaluation in recent years.

{\textbf{CodeBLEU}}~\cite{ren2020codebleumethodautomaticevaluation} addresses the inadequacy of natural language evaluators in capturing code syntax and semantics. It provides a similarity score using the weighted average of four different components: the standard BLEU score \cite{10.3115/1073083.1073135}, a weighted n-gram match emphasizing programming language keywords, a syntactic AST match, and a semantic data-flow match. Since its introduction, it has remained the most widely adopted static metric for code evaluation.

{\textbf{CrystalBLEU}}~\cite{10.1145/3551349.3556903}, was proposed with the claim of achieving improved \textit{distinguishability} (detailed in Section \ref{distinguishability}) over BLEU \cite{10.3115/1073083.1073135} and CodeBLEU. Since source code can share trivial n-grams despite differing in functionality, CrystalBLEU provides a similarity score excluding these \textit{trivially shared n-grams}.

{\textbf{CodeBERTScore}} uses language models pretrained on code to create embeddings for both the reference code and the generated code and then computes a semantics-aware similarity score by comparing the embeddings. CodeBERTScore has shown stronger correlation with both human preference and functional correctness compared to other competing metrics \cite{zhou2023codebertscore}.

{\textbf{CodeScore}}~\cite{10.1145/3695991} is the most recently proposed metric among those we evaluate and has not yet been audited. It specifically targets the \textit{surface bias} in CEMs by training a large language model (LLM) to estimate functional correctness from execution results.

\subsection{Evaluating the effect of \textit{Surface Bias} on CEMs}
While there is a common notion among the community that reference-based static CEMs are biased by surface similarity \cite{EVTIKHIEV2023111741, naik2024limitationsembeddingbasedmethods}, to the best of our knowledge, we are the first to quantify the effect size of \textit{surface bias} on these metrics. We evaluate the extent to which these metrics \textit{fail to reflect true semantic similarity} in two independent ways - first, by measuring how strongly the scores provided by the metrics correlate with surface similarity, and second, by evaluating the \textit{distinguishability} (detailed in Section \ref{distinguishability}) of the metrics in the presence of \textit{surface bias}. 
\subsubsection{Measuring the Correlation between Surface Similarity and Metrics Output}
To account for the surface similarity between two codes in a pair, we consider both textual and structural similarity.
\textbf{Inverse Edit Distance, $Sim_{Edit}$ :} Treats the codes as strings and captures their textual similarity by leveraging the inverse of edit distance. Here, edit distance is the minimum number of insertions, deletions, or substitutions needed to transform one code into the other in a pair \cite{edit1, edit2}. 
\textbf{AST similarity, $Sim_{AST}$: } Assesses the structural similarity between two codes in a pair by comparing their Abstract Syntax Trees (AST) \cite{ast}. 

Unless otherwise mentioned, we calculate the surface similarity between two codes $c_{1}$ and $c_{2}$ as:
\begin{equation}
\text{SurfaceSim}(c_1, c_2) = \frac{1}{2} \left[  \text{Sim}_{\mathrm{Edit}}(c_1, c_2) + \text{Sim}_{\mathrm{AST}}(c_1, c_2) \right]
\label{eq:surface_similarity}
\end{equation}
We quantify the correlation between \textit{SurfaceSim} and the scores from the four metrics in Section \ref{RQ1} and statistically evaluate the significance with which this correlation holds. 
\par While in this paper, we use an unweighted mean to calculate \textit{SurfaceSim}, future work could adjust the weights in  Eq. \ref{eq:surface_similarity} to find combinations that reveal greater \textit{surface bias}. 
\subsubsection{Evaluating Distinguishability in the Presence of Surface Bias } CrystalBLEU \cite{10.1145/3551349.3556903} formulates a meta-metric, \textit{distinguishability}, to evaluate a CEM's ability to isolate functionally equivalent code pairs from the non-equivalent ones.
\begin{definition}[Distinguishability{~\cite[Def. 3.1]{10.1145/3551349.3556903}}]
The distinguishability of a metric $m$ is
\begin{equation}
d = \frac{m(\text{Pairs}_{\text{intra}})}{m(\text{Pairs}_{\text{inter}})}.
\label{eq:distinguishability}
\end{equation} 
where $\mathcal{C}$ is the set of all code snippets, partitioned into disjoint equivalence classes 
$\mathcal{C}_1,\ldots,\mathcal{C}_n$ and $\text{Pairs}_{\text{intra}}=\{(\mathcal{C}_i\setminus\{c\},c)\mid c\in\mathcal{C}_i\}$, $\text{Pairs}_{\text{inter}}=\{(\mathcal{C}_j,c)\mid c\in\mathcal{C}_i,\, i\neq j\}$ 
for $i,j\in\{1,\ldots,n\}$.
\end{definition}

We adapt this definition to the pairwise setting by defining
$
\text{Pairs}_{\text{intra}} = \{ (c_1, c_2) \mid c_1, c_2 \in \mathcal{C}_i \ \text{for some} \ i \}
$ and $
\text{Pairs}_{\text{inter}} = \{ (c_1, c_2) \mid c_1 \in \mathcal{C}_i, \ c_2 \in \mathcal{C}_j, \ i \neq j \}
$.
Let
$m : (C \times C) \to \mathbb{R}_{\geq 0}$
be a CEM that assigns a non-negative real-valued similarity score to a code pair. 
We then define the distinguishability of a CEM, $m$, as:
\begin{equation}
d = \frac{\frac{1}{|\text{Pairs}_{\text{intra}}|} \sum\limits_{(c_1,c_2) \in \text{Pairs}_{\text{intra}}} m(c_1, c_2)}
         {\frac{1}{|\text{Pairs}_{\text{inter}}|} \sum\limits_{(c_1,c_2) \in \text{Pairs}_{\text{inter}}} m(c_1, c_2)}
\label{eq:distinguishability}
\end{equation}
A value of $d>1$ indicates $m$ assigns higher scores  
to functionally equivalent pairs than to the non-equivalent ones. While CrystalBLEU reports higher \textit{distinguishability} compared to other metrics, we hypothesize that this claim holds only when the non-equivalent pairs differ substantially in appearance. To evaluate our hypothesis, in Section \ref{RQ2}, we replace the non-equivalent pairs used by Eghbali et al. in CrystalBLEU with more similar-looking non-equivalent pairs and analyze how this affects the \textit{distinguishability} of CEMs. 
\label{distinguishability}

\subsection{Benchmark LoCaL - Looks Can Lie}
\label{LoCal_consttruction}
We define two critical evaluation regions for a CEM - \textbf{SFD (Similar Form, Different Semantics)}, which includes code pairs with high surface similarity but low functional similarity, and \textbf{DFS (Different Form, Similar Semantics)}, which provides for code pairs with high functional similarity but low surface similarity.  To address the lack of benchmarks that have adequate coverage in the SFD and DFS regions, we propose our benchmark, LoCaL (Looks Can Lie), which specifically targets generating code pairs within the SFD and DFS regions and measures their similarity scores using a differential fuzzing pipeline.  
\subsubsection{Data Selection and Preparation} We build LoCaL on the test splits of four widely used datasets- HumanEval \cite{chen2021evaluatinglargelanguagemodels} and MBPP \cite{austin2021programsynthesislargelanguage} for function-level code pairs and APPS \cite{hendrycks2021measuring} and PIE \cite{shypula2024learning} for program-level code pairs. HumanEval, MBPP, and APPS are code generation benchmarks, whereas PIE evaluates code optimization. We particularly include PIE since it has not been used in training CodeScore, one of the metrics under test, and hence, remains unseen for the metric. For every task, each of these benchmarks provides at least one reference solution and test cases to evaluate any generated code against the reference. 
PIE provides a slow implementation and a functionally equivalent, faster version for each task. 
Each pair in LoCaL requires a substantial number of test cases to be executed during evaluation, leading to increased runtime. We, therefore, sort the entries in the original benchmark from longest to shortest (based on lines of code) and then take the top 100 samples per dataset, except HumanEval. We take HumanEval in full since it has only 164 entries. Longer codes are prioritized as they can be modified in more ways to generate variants. 
Table \ref{tab:datasets} summarizes the datasets used to construct LoCaL. 
\begin{table}[t]
\centering
\caption{Summary of datasets used to construct LoCaL. ``\# Ori.'' denotes the number of original problems in the benchmark, and ``k'' denotes the number of problems selected in LoCaL.}
\vspace{-3mm}
\label{tab:datasets}
\renewcommand{\arraystretch}{1.2}
\small
\begin{tabularx}{0.9\linewidth}{l c c c c c}
\toprule
\textbf{Benchmark} & \textbf{\# Ori.} & \textbf{k} & \textbf{\# Test Cases/Problem} & \textbf{NL description?} & \textbf{Level}\\
\hline
HumanEval & 164  & 164      & 7.7 & Yes & Function \\
MBPP      & 974  &  100       & 3.0 & Yes & Function \\
APPS      & 5000 &  100      & 21.2 & Yes & Program  \\
PIE & 1000 & 100\makebox[0pt][l]{\textsuperscript{*}} & 104 & No & Program \\
\bottomrule
\end{tabularx}
\par\vspace{2pt}
\footnotesize{$^*$ PIE already has optimized variants, only mutated variants are generated from $k$ samples.}
\end{table}

\par Since the source benchmarks vary in their original format, we reduce each benchmark to four particulars: a unique \textbf{task ID} for each entry, the \textbf{natural language description} of the problem, when available, the \textbf{reference solution}, and the \textbf{test cases} to evaluate any other solution for that task ID.

\begin{figure}[t]
  \centering
  \includegraphics[width=0.8\linewidth]{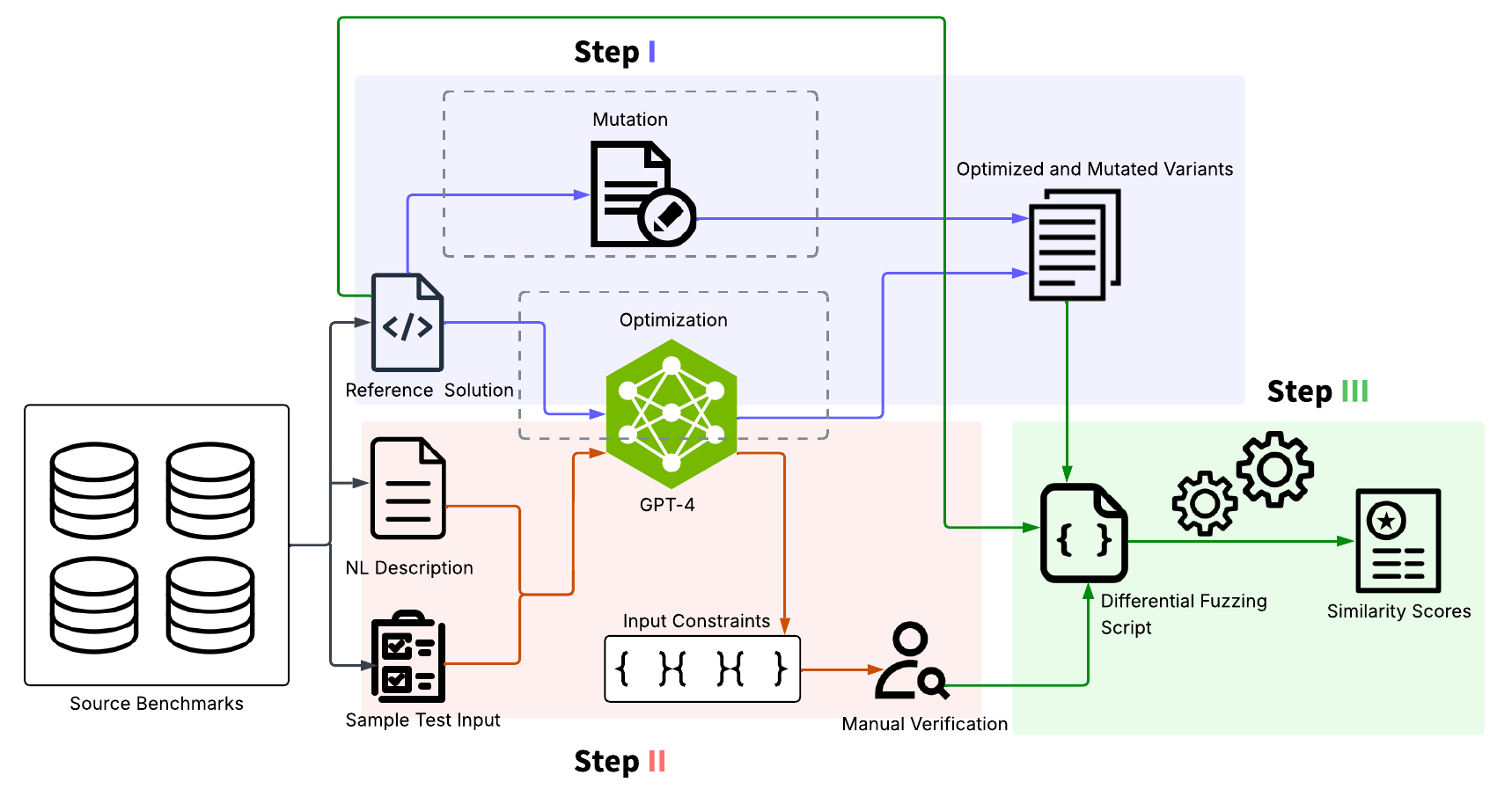} 
  \vspace{-5mm}
  \caption{The LoCaL construction pipeline}
  \label{fig:local-pipeline}
\end{figure}

\subsubsection{Benchmark Construction}
The construction of LoCaL can be described in three steps, as depicted in Figure~\ref{fig:local-pipeline}: \Romannum{1}) Optimized and Mutated variant generation, \Romannum{2}) Test Input Constraints Definition, and \Romannum{3}) Differential Fuzzing. \\

\textbf{Step \Romannum{1} : Optimized and Mutated Variant Generation.} We focus on generating code pairs that have a sharp contrast between their surface similarity and functional similarity to effectively hit the SFD and DFS regions. To create code pairs that alter surface form while preserving the semantics, we opt for performance optimization. Unlike previous work that applies arbitrary code edits to introduce structural changes \cite{Nacl}, we ask GPT-4 \cite{openai2024gpt4technicalreport}, a large language model (LLM), to optimize each reference solution using one or more of these five strategies: 1) algorithmic \cite{algo}, 2) data structures \cite{ds}, 3) cold path \cite{hot-cold}, 4) hot path \cite{hot-cold}, and 5) memoization \cite{Michie1968MemoFA}, thereby reflecting structural changes that are likely to occur in real development scenarios. 

However, we acknowledge that not all implementations can be optimized and hence ask the LLM to return ``None" as a strategy if the code is not optimizable. We then record the optimized code, along with all applied strategies, as an optimized variant of the reference solution in LoCaL. For altering semantics while preserving most of the surface similarity, we modify MutPy \cite{MutPy}, a Python mutation testing tool, to preserve all the mutants it generates. We run MutPy with a dummy test suite, forcing it to work only as a mutant generation engine. It applies operators like AOD (arithmetic operator deletion) and ZIL (zero iteration loop) to introduce small changes to the source code and produces numerous semantically altered but similar-looking mutants of the reference solution, all stored as mutated variants along with the mutation operator applied. Since CEMs have been applied to tasks like code translation, code optimization, and automated bug repair \cite{Zhu_Suresh_Reddy_2022, 10.1145/3691620.3695537, 11024270, ops}, where it's extremely important to capture code semantics going beyond lexical overlap, we think optimization and mutation are justified choices to assess the reliability of these metrics. For the rest of the paper, we refer to the reference solution as $code_{ori}$ and its variants as $code_{var}$. \\ 

\begin{figure}[t]
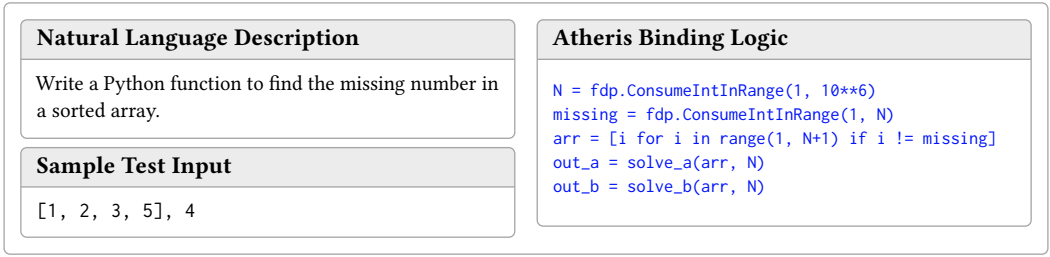

\centering
\setlength{\columnsep}{8pt}
\begin{tcolorbox}[mybox, width=\linewidth, title={}]
\begin{multicols}{2}

\begin{tcolorbox}[mybox, title=Natural Language Description]
\footnotesize
Write a Python function to find the missing number in a sorted array.
\end{tcolorbox}

\begin{tcolorbox}[mybox, title=Sample Test Input]
\footnotesize
\texttt{[1, 2, 3, 5], 4}
\end{tcolorbox}
\columnbreak
\begin{tcolorbox}[mybox, title=Atheris Binding Logic]
\lstset{style=ath}
\begin{lstlisting}
N = fdp.ConsumeIntInRange(1, 10**6)
missing = fdp.ConsumeIntInRange(1, N)
arr = [i for i in range(1, N+1) if i != missing]
out_a = solve_a(arr, N)
out_b = solve_b(arr, N)
\end{lstlisting}
\end{tcolorbox}
\end{multicols}
\end{tcolorbox}
\vspace{-3mm}
\caption{Example of test input constraint definition for MBPP task 34}
\label{fig:input_constraints_example}
\end{figure}

\textbf{Step \Romannum{2}: Test Input Constraints Definition.} Once we have the $(code_{ori}, code_{var})$ pairs, we generate a large number of test inputs for each pair using Atheris \cite{atheris}, a byte-level fuzzer. Since Atheris mutates bytes without domain awareness, a key challenge here is ensuring that the generated test inputs are valid and relevant. For example, as illustrated in Figure \ref{fig:input_constraints_example}, if a problem requires two arguments, a sorted array and an integer representing its size, the fuzzer must only generate the array elements in sorted order and set the second argument to the array's size, rather than any random integer. We ensure this in two steps.
First, we prompt (few-shot) GPT-4 with the natural language task description and an example test input to produce Atheris-compatible input constraints. For the PIE dataset \cite{shypula2024learning}, which lacks a natural language description, we utilize an example test input and the reference implementation. 
Second, we manually verify all GPT-generated input binding logic for each problem to ensure it is correct, relevant, and preserves all relationships between the input parameters. 

\textbf{Step \Romannum{3} : Differential Fuzzing.}  
Algorithm  \ref{alg:dfsim_full} presents the final step of our pipeline, in which a differential fuzzing script is prepared for each $(code_{ori}, code_{var})$ pair.

For function-level codes, the differential fuzzing script defines the $code_{ori}$ as $\mathit{func}\_{a}$ and $code_{var}$ as $\mathit{func}\_{b}$. Using Atheris, we generate 2000 test inputs (constrained according to Step II) and pass them to both functions. For program-level codes, $code_{ori}$ and $code_{var}$ are stored in separate Python files. This time, we generate 1000 constrained test inputs due to the higher runtime cost and run both Python files on these inputs. We exclude time-outs and measure the functional similarity score for a pair as the proportion of inputs that produce identical outputs in both implementations. To reduce randomness, each differential fuzzing script is run five times, and the final differential fuzzing score, $\mathbf{df_{score}}$ is the average across these runs. 




\begin{algorithm}[t]
\footnotesize
\SetAlgoNlRelativeSize{-1}
\SetAlCapSkip{0.5em}
\caption{Differential Fuzzing Similarity for a Code Pair in LoCaL}
\label{alg:dfsim_full}
\KwIn{$code_{ori},\, code_{var}$; input domain $\mathcal{D}$; number of tests $N$; number of repetitions $R$, time budget $T$}
\KwOut{$\mathrm{Sim}\in[0,1]$ or $-1$ to indicate that all tests timed out.}

$S \gets \varnothing$\;
\For{$i \gets 1$ \KwTo $R$}{
 $t_0 \gets \mathrm{now}()$\;
  Generate $\{x_j\}_{j=1}^N$ with Atheris over $\mathcal{D}$\;
  matches $\gets 0$; $\mathit{timedOut} \gets \mathrm{false}$\;
  \For{$j \gets 1$ \KwTo $N$}{
   \If{$\mathrm{now}() - t_0 > T$}{$\mathit{timedOut} \gets \mathrm{true}$; \textbf{break}}
    $(y^{ori}, y^{var}) \gets (code_{ori}(x_j),\, code_{var}(x_j))$\;
    \If{$y^{ori} = y^{var}$}{matches$++$\;}
  }
  \If{$\mathit{timedOut}$}{\textbf{continue}}
  $S \gets S \cup \left\{\frac{\text{matches}}{N}\right\}$\;
}
\lIf{$|S|=0$}{\Return $-1$}
\Return $\frac{1}{|S|}\sum_{s\in S} s$\;
\end{algorithm}

\subsubsection{Benchmark Characteristics}

Excluding timeouts, LoCaL contains 3117 Python code pairs, each labeled with a functional similarity score quantified by differential fuzzing. In addition, each pair includes its surface similarity and the optimization or mutation strategy used to derive it. Optimized variants for PIE \cite{shypula2024learning} do not include a strategy since they were already optimized in the source benchmark. The statistics of LoCaL are given in Table \ref{tab:local_stats}. Several factors explain why optimized variants are fewer than mutated variants in LoCaL. First, optimization opportunities are limited - not all codes can be optimized, and even for the optimizable implementations, we obtain very few variants, often just 1. MutPy, on the other hand, can generate many mutants for a single reference implementation. Moreover, GPT-generated optimizations fail to compile more often than the MutPy mutants. As shown in Table \ref{tab:local_stats}, mutations exhibit very high surface similarity on average since they result from small, localized edits in the reference implementation.

Optimizations, on the other hand, show 0.644 surface similarity on average, indicating generative models make non-trivial structural changes in a code when asked to optimize.
\begin{table}[t]
\centering
\caption{Statistics of LoCaL}
\vspace{-3mm}
\small
\label{tab:local_stats}
\begin{tabular}{lccccc}
\toprule
\textbf{Variant Type} & \textbf{Method-Level} & \textbf{Program-Level} & \textbf{Total} &\textbf{Avg. SurfaceSim} & \textbf{Avg. $\mathbf{df_{score}}$}\\
\midrule
Optimized & 200 & 265  & 465 & 0.644 & 0.985 \\
Mutated     & 1148 & 1504 & 2652& 0.922 & 0.093\\
\bottomrule
\end{tabular}
\end{table}

\subsection{Defining SFD and DFS regions}
\label{sec:SFD-DFS}
To test the coverage of existing benchmarks and LoCaL within the SFD and DFS regions, we mathematically define the thresholds that bound these regions.

\begin{figure}[t]
  \centering
  \includegraphics[width=0.70\linewidth]{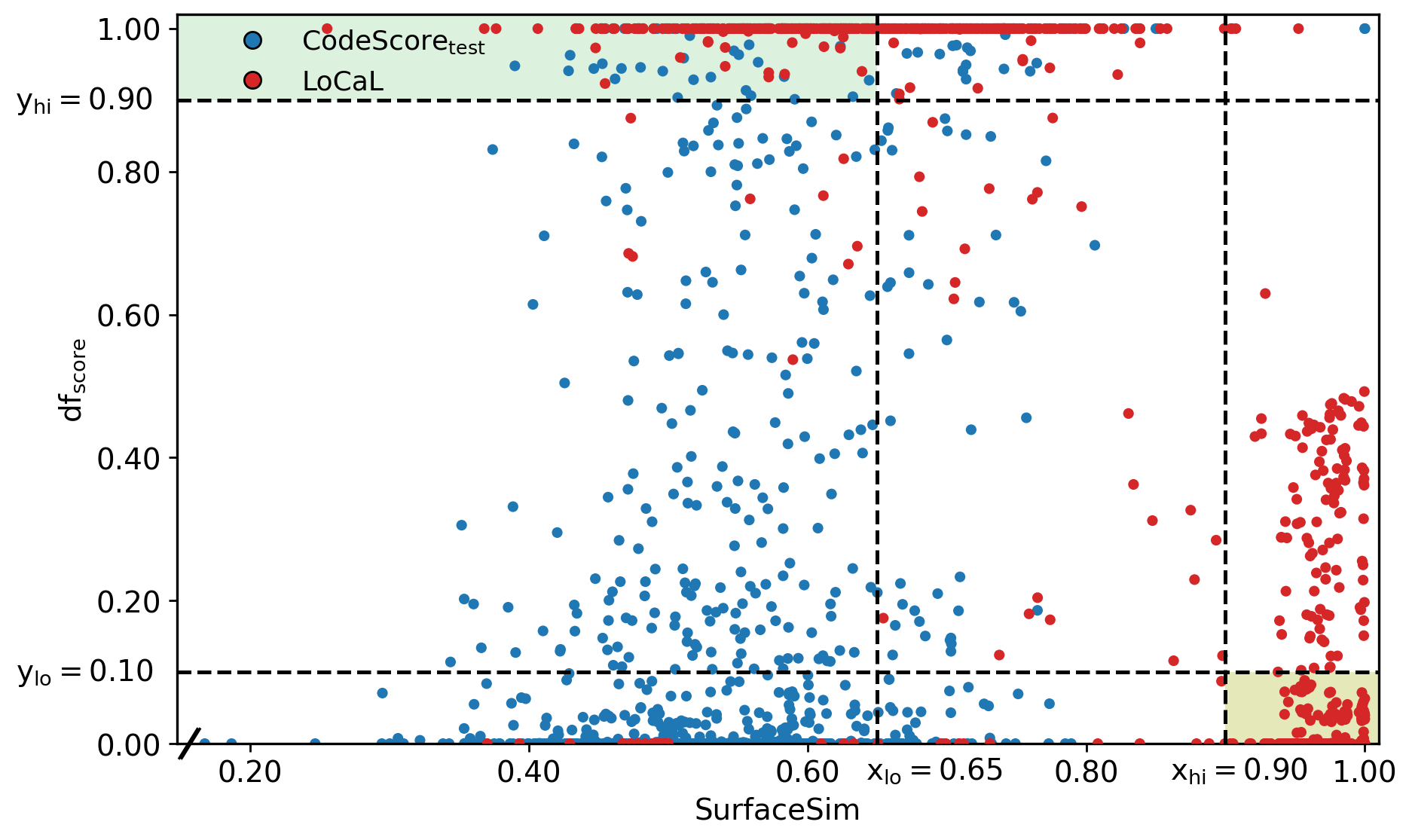} 
  \vspace{-4mm}
  \caption{Data distribution across regions: LoCaL vs Baseline. DFS (green), SFD (yellow), and Control (unshaded) regions are defined by the thresholds calculated using Alg. \ref{alg:mse_gap}.}

  \label{fig:RQ3_results}
\end{figure}

In the SurfaceSim vs. $df_{score}$ space (Fig. \ref{fig:RQ3_results}), \textbf{SFD (Similar Form, Different Semantics)} denotes the yellow bottom-right corner where surface similarity between two codes is greater than their functional similarity. The green top-left corner, where functional similarity exceeds surface similarity, defines \textbf{DFS (Different Form, Similar Semantics)}. We define the remainder of the space as \textbf{Control}.

\begin{definition}[SFD, DFS, and Control region]
Let $x\in[0,1]$ denote SurfaceSim, $y\in[0,1]$ denote $df_{\mathrm{score}}$ and $x=x_{lo}, x=x_{hi}, y=y_{lo}, y=y_{hi}$ be four axis-aligned lines where $x_{lo}<x_{hi}$ and $y_{lo}<y_{hi}$. We then define, 
\begin{equation}\label{eq:sfd-dfs-control}
\begin{aligned}
\mathrm{SFD} &:= \{(x,y)\in[0,1]^2 \mid x \ge x_{hi},\; y \le y_{lo}\},\\
\mathrm{DFS} &:= \{(x,y)\in[0,1]^2 \mid x \le x_{lo},\; y \ge y_{hi}\},\\
\mathrm{Control} &:= [0,1]^2 \setminus \bigl(\mathrm{SFD}\cup \mathrm{DFS}\bigr).
\end{aligned}
\end{equation}
\end{definition}
We use Algorithm \ref{alg:mse_gap} to compute thresholds $(x_{lo}, x_{hi}, y_{lo},$ and $y_{hi})$ for our experiments, that maximize the difference in prediction error between points inside and outside the induced boundary. We think this is a reasonable heuristic to isolate the ``corner" regions where CEMs are likely to struggle more than the rest of the space. The algorithm takes a dataset, $\mathcal{D}$, a step size, $\Delta$, and a set of CEMs, $\mathcal{M}$ as inputs. Each row in $\mathcal{D}$ has two codes, their SurfaceSim (Eq. \ref{eq:surface_similarity}), their $df_{score}$, and the scores given to the code pair by each metric in $\mathcal{M}$. Using the step size, $\Delta$, the algorithm scans a grid of candidate threshold tuples $(x_{lo}, x_{hi}, y_{lo}, y_{hi})$ and partitions the data into SFD, DFS, and Control regions as defined in Eq. \ref{eq:sfd-dfs-control}. Then, for every metric, it computes the mean absolute error (MAE) in each region with respect to the ground truth, $df_{score}$. Next, it calculates two gaps, $MAE_{DFS} - MAE_{Control}$ and $MAE_{SFD} - MAE_{Control}$. Finally, it averages the gaps across all metrics and returns the threshold tuple that maximizes this average \textit{Control vs. Corner} gap. 
\begin{algorithm}[t]
\footnotesize
\SetAlgoNlRelativeSize{-1}
\SetAlCapSkip{0.5em}
\caption{Threshold Selection for SFD and DFS}
\label{alg:mse_gap}
\KwIn{Dataset $\mathcal{D}=\{(x_r,y_r,\{m_k(r)\}_{k\in\mathcal{M}})\}$ where for each $r\in\mathcal{D}$, $x_r$ = SurfaceSim, $y_r=df_{\mathrm{score}}$, and $x_r,y_r\in[0,1]$; step size $\Delta$; $\mathcal{M}$ = set of CEMs.\\}

\KwOut{$(x_{\mathrm{lo}},x_{\mathrm{hi}},y_{\mathrm{lo}},y_{\mathrm{hi}})$.}

$\text{best}\gets -\infty$\;

\For{$x_1$ from $0$ to $1$ \textbf{step} $\Delta$}{
\For{$x_2$ from $0$ to $1$ \textbf{step} $\Delta$}{
\For{$y_1$ from $0$ to $1$ \textbf{step} $\Delta$}{
\For{$y_2$ from $0$ to $1$ \textbf{step} $\Delta$}{
  \If{$\neg(x_1<x_2\ \land\ y_1<y_2)$}{\textbf{continue}}
  \DontPrintSemicolon
  Partition $\mathcal{D}$:\;     
  \Indp
  \(                          
  \begin{aligned}[t]
  \mathrm{DFS}  &= \{\, r:\ x_r \le x_1,\ y_r \ge y_2 \,\},\\
  \mathrm{SFD}  &= \{\, r:\ x_r \ge x_2,\ y_r \le y_1 \,\},\\
  \mathrm{CTRL} &= \mathcal{D}\setminus(\mathrm{DFS}\cup\mathrm{SFD})
  \end{aligned}
  \)\;                        
  \Indm
  \For{$k\in\mathcal{M}$}{
    $\mathrm{MAE}_Z(k)=\frac{1}{|Z|}\sum_{r\in Z}(y_r-m_k(r))^2$ for $Z\in\{\mathrm{DFS},\mathrm{SFD},\mathrm{CTRL}\}$\;
    $g_{\mathrm{DFS}}(k)=\mathrm{MAE}_{\mathrm{DFS}}(k)-\mathrm{MAE}_{\mathrm{CTRL}}(k)$\;
    $g_{\mathrm{SFD}}(k)=\mathrm{MAE}_{\mathrm{SFD}}(k)-\mathrm{MAE}_{\mathrm{CTRL}}(k)$\;
  }
  $\mathrm{obj}=\displaystyle \mathrm{mean}_{k\in\mathcal{M}}\Bigl(\mathrm{mean}\{g_{\mathrm{DFS}}(k),\,g_{\mathrm{SFD}}(k)\}\Bigr)$\;
  \If{$\mathrm{obj}>\text{best}$}{
    $\text{best}\gets \mathrm{obj}$;\ 
    $(x_{\mathrm{lo}},x_{\mathrm{hi}},y_{\mathrm{lo}},y_{\mathrm{hi}})\gets(x_1,x_2,y_1,y_2)$\;
  }
}}}}

\Return $(x_{\mathrm{lo}},x_{\mathrm{hi}},y_{\mathrm{lo}},y_{\mathrm{hi}})$\;
\end{algorithm}

\section{Experimental Study}
\label{sec:experiments}
To quantify the impact of \textit{surface bias} on CEMs and to evaluate the effectiveness of LoCaL, we answer the following \textit{five} research questions :\\
\noindent\textbf{RQ1:} How strong is the correlation between surface similarity and CEMs?\\
\noindent\textbf{RQ2:} How is the distinguishability of CEMs affected by the presence of \textit{surface bias}?\\
\noindent\textbf{RQ3:}  How adequately does LoCaL cover the SFD and DFS regions compared to existing work?\\
\noindent\textbf{RQ4:} How effective is the benchmark, LoCaL, in revealing the limitations of CEMs?\\
\noindent\textbf{RQ5:} Does integrating LoCaL samples in training improve CodeScore's performance?

\subsection{Baseline}
\label{baselines}
Dong et al. \cite{10.1145/3695991} built three inference datasets, named APPS-Eval, MBPP-Eval, and HE-Eval, all of which provide continuous functional similarity scores between two Python implementations. While we adopt their code pairs, we do not use their similarity scores to ensure a common scale with LoCaL. Instead, we rerun the code pairs through our differential-fuzzing pipeline to obtain $df_{score}$. To manage runtime, we randomly sample 500 entries from APPS-Eval (program-level) and 500 samples from MBPP-Eval (method-level). Excluding time-outs, this yields 854 samples in total, which we use as our baseline, \textbf{CodeScore\textsubscript{test}}. 
   
\subsection{Evaluation Metric} We compute the prediction error for a CEM, $k$, on a dataset, $\mathcal{D}$, using mean absolute error (MAE). Our metric is
$
\mathrm{MAE}_k
= \frac{1}{|\mathcal{D}|}\sum_{r\in\mathcal{D}} \bigl| m_k(r) - y_r \bigr| 
\label{eq:mae_k}
$, 
where $r$ is a code pair, $m_{k}(r)$ is the score that CEM, $k$ assigns to $r$, and $y_r$ is the ground truth functional similarity score for $r$ in the dataset, $\mathcal{D}$. We prefer MAE since it provides a clear, uniform estimate of the absolute deviation from the ground truth, unlike mean squared error (MSE), which over-punishes large, rare errors. 
\label{metrics}
\subsection{RQ1: Correlation Between Surface Similarity and CEMs}
\label{RQ1}
\subsubsection{Experimental Setup} In this experiment, we test if the CEMs show any correlation with surface similarity when functional similarity remains constant. We use the PIE \cite{shypula2024learning} dataset, which contains 1000 pairs of \textit{slow-fast} equivalents (functional similarity 1.0). For each pair in the dataset, we calculate its SurfaceSim (Eq. \ref{eq:surface_similarity}), and the scores from CodeBLEU, CrystalBLEU, CodeBERTScore, and CodeScore. To assess the correlation between a CEM and surface similarity, we rely on \textbf{Spearman’s rank correlation} \cite{spearman1904}, $\boldsymbol{r}_{\rho}$. Since our variables are not normally distributed (validated using the Shapiro-Wilk normality test \cite{10.1093/biomet/52.3-4.591}), the Spearman coefficient is more appropriate than the Kendall \cite{665905b2-6123-3642-832e-05dbc1f48979} or Pearson coefficient \cite{pearson1896} in this setting. 

\begin{table}[t]
\centering
\caption{Spearman correlation ($\mathbf{r}_{\rho}$) between surface similarity and CEMs with functional similarity of 1.0.}
\vspace{-2mm}
\label{tab:spearman_corr}
\renewcommand{\arraystretch}{1.2}
\small
\begin{tabular}{l c c c}
\toprule
\textbf{Metric} & \textbf{n} & $\mathbf{r}_{\rho}$ & \textbf{p-value} \\
\hline
CodeBLEU        & 1000 & 0.8732 & $9.61 \times 10^{-314}$ \\
CrystalBLEU     & 1000 & 0.5886 & $2.76 \times 10^{-94}$ \\
CodeBERTScore   & 1000 & 0.8811 & $4.94 \times 10^{-324}$ \\
CodeScore       & 1000 & 0.3481 & $7.26 \times 10^{-30}$ \\
\bottomrule
\end{tabular}
\end{table}

\subsubsection{Results and Analysis}
From Table \ref{tab:spearman_corr}, we find that all four CEMs have a positive correlation with surface similarity, whereas in the ideal case, they should be invariant and hence show zero correlation. For CodeBLEU and CodeBERTScore, the correlation is \textbf{very strong} at $\approx$ 0.88. This clearly implies that the metrics are measuring the wrong dimensions of similarity. CrystalBLEU also shows a \textbf{moderately strong} correlation of $\approx$ 0.59, while CodeScore, though \textbf{weak to moderate} at $\approx$ 0.35, still demonstrates an undesirable dependence on surface similarity. Interestingly, CrystalBLEU often returns zero due to its $k$ shared n-gram filtering, which eliminates almost all overlaps for some pairs. With fewer zeros, the correlation for CrystalBLEU would likely be higher. Although we follow prior work \cite{naik2024limitationsembeddingbasedmethods} and use $k=50$ while using CrystalBLEU in our experiments, determining a Python-appropriate cut-off for CrystalBLEU requires further study. We also observe that CodeScore produces highly variable scores on unseen data, which suggests that there might be other confounding factors, along with surface similarity, contributing to its poor performance. 

With p-values ranging from $10^{-30}$ to $10^{-324}$ across the CEMs, the positive correlations shown in Table \ref{tab:spearman_corr} hold with significance. 

\begin{rqbox}
\textbf{Findings of RQ1:} We observe a \textbf{positive correlation} between every CEM and surface similarity, with CodeBLEU and CodeBERTScore having the strongest positive correlation. Hence, the scores assigned by the CEMs to a code pair are predominantly influenced by the surface-level overlap between the codes in the pair.  
\end{rqbox}
\subsection{RQ2: Distinguishability Analysis}
\label{RQ2}
\subsubsection{Experimental Setup} While Eghbali et al. use a Java dataset aggregated from \textit{ShareCode.io} \cite{sharecode} for distinguishability analysis \cite{10.1145/3551349.3556903}, we opt for a Python dataset since CodeScore can evaluate Python pairs only. 

Following \cite{10.1145/3551349.3556903}, we build a dataset, $ds_{orig}$, with 100 functionally equivalent and 100 functionally non-equivalent pairs randomly drawn from one of our source benchmarks, PIE \cite{shypula2024learning}. PIE already provides for pairs of slow and fast equivalents. To create the non-equivalent pairs in $ds_{orig}$, we replicate the cross-pairing strategy used in \cite{10.1145/3551349.3556903}, i.e., taking the slow solution to one problem and pairing it with the fast solution of another, thereby forming pairs that cannot be functionally equivalent. We create a second dataset, $ds_{replaced}$, by replacing the non-equivalent pairs in $ds_{orig}$ with 100 new pairs that are more similar-looking, but still functionally non-equivalent. The new pairs are generated by mutating slow implementations with MutPy, the same way mutated variants are generated in LoCaL (detailed in Section \ref{LoCal_consttruction}). We then calculate the \textit{distinguishability} of the four CEMS on both $ds_{orig}$ and $ds_{replaced}$  using Eq. \ref{eq:distinguishability}. 

\par We enforce that all new pairs in $ds_{replaced}$ have $df_{score}=0$ to ensure reliable non-equivalence. Since many mutants do not reach a strict 0, and confirming this with differential fuzzing is expensive, we limit the sample size to 100 for this experiment. However, to account for any noise arising from random sampling, we repeat the entire dataset construction till the distinguishability calculation process 10 times and report the mean$(\pm$standard deviation)
in Table \ref{tab:RQ2_results}. The surface similarity (Eq. \ref{eq:surface_similarity}) for equivalent pairs is $\mathbf{0.76(\pm0.02)}$ in both datasets, while non-equivalents yield $\mathbf{0.49(\pm0.01)}$ in $ds_{orig}$ and $\mathbf{0.89(\pm0.02)}$ in $ds_{replaced}$.

\begin{table}[t]
\centering
\caption{Impact of \textit{surface bias} on the \textit{distinguishability}, $\mathbf{d}$ of CEMs. EQ Score is the score assigned to the equivalent pairs, and NEQ Score is the score assigned to the non-equivalent pairs by each CEM. Equivalent pairs are shared across datasets and hence, shown once.}
\vspace{-2mm}
\label{tab:RQ2_results}
\small
\setlength{\tabcolsep}{6pt}
\renewcommand{\arraystretch}{1.3}
\begin{tabular}{l|c|c|c|c|c}
\toprule
\multirow{2}{*}{\textbf{CEM}} &
\multirow{2}{*}{\textbf{EQ Score}} &
\multicolumn{2}{c|}{\textbf{ds\textsubscript{orig}}} &
\multicolumn{2}{c}{\textbf{ds\textsubscript{replaced}}} \\
\cmidrule(lr){3-4}\cmidrule(lr){5-6}
& & \textbf{NEQ Score} & $\mathbf{d}$ & \textbf{NEQ Score} & $\mathbf{d}$ \\
\midrule
CodeBLEU        & $0.49(\pm0.03)$ & $0.17(\pm0.01)$ & $2.97(\pm0.21)$ & $0.97(\pm0.00)$ & $0.51(\pm0.03)$\\
CrystalBLEU     & $0.39(\pm0.03)$ & $0.05(\pm0.00)$ & $8.60(\pm1.40)$ & $0.98(\pm0.00)$ & $0.40(\pm0.03)$\\
CodeBERTScore   & $0.89(\pm0.01)$ & $0.71(\pm0.01)$ & $1.25(\pm0.02)$ & $0.99(\pm0.00)$ & $0.90(\pm0.01)$ \\
CodeScore       & $0.59(\pm0.04)$ & $0.45(\pm0.05)$ & $1.33(\pm0.15)$ & $0.98(\pm0.01)$ & $0.60(\pm0.04)$ \\
\bottomrule
\end{tabular}
\end{table}

\subsubsection{Results and Analysis} As shown in Table \ref{tab:RQ2_results}, all four CEMs have a distinguishability over 1 on $ds_{orig}$, whereas every metric's distinguishability collapses below 1 on $ds_{replaced}$, with CrystalBLEU, CodeBLEU, CodeScore, and CodeBERTScore showing a drop of \textbf{95.3\%}, \textbf{82.8\%}, \textbf{54.9\%}, and \textbf{28.0\%}, respectively. 
This happens because in $ds_{orig}$, the non-equivalent pairs have less surface similarity than the equivalent pairs, so the CEMs assign lower scores to them. But, when in ds\textsubscript{replaced}, the non-equivalent pairs look more alike than the equivalent ones, all four CEMs get confused and provide higher scores to the non-equivalent pairs instead, although \textbf{1000 test inputs} validate their non-equivalence. 

On $ds_{orig}$, CrystalBLEU reports an impressive distinguishability of 8.60, which validates the authors' claim of improved distinguishability over other metrics. However, we find a serious flaw in their experimental design. Since in $ds_{orig}$, the non-equivalent pairs are taken by cross-pairing solutions to different problems, the only probable overlap in a pair is the basic but frequently occurring Python syntax or keywords such as \textit{def}, \textit{for}, \textit{in}, etc. As CrystalBLEU filters out these trivially shared n-grams, minimal surface similarity remains, allowing it to be effective in comparing solutions that belong to different problems. However, it is clear from Table \ref{tab:RQ2_results} that when the non-equivalent pairs have richer surface overlap than trivially shared n-grams. CrystalBLEU, in fact, performs the worst in detecting equivalence. 
\par A distinguishability near 1 on $ds_{replaced}$ does not mean CodeBERTScore separates functional classes better. This is only because the metric assigns high scores to both equivalent and non-equivalent pairs; for example, it provides an average score of 0.71 to even dissimilar non-equivalent pairs (Table \ref{tab:RQ2_results}). This forces the mean scores for equivalent and non-equivalent pairs to be close, and hence, their ratio (\textit{distinguishability}) always stays near 1, with very little room to vary. 
\begin{rqbox}
\textbf{Findings of RQ2: }All four CEMs lose their ability to distinguish between functionally equivalent and non-equivalent pairs in the presence of \textit{surface bias}. Experiments prove that the higher distinguishability claim of CrystaLBLEU is only valid when functional similarity closely matches surface similarity. In scenarios where they diverge, the claim does not hold. 
\end{rqbox}

\subsection{RQ3: Coverage Analysis of the SFD and DFS regions} 
\label{RQ3}
\subsubsection{Experimental Setup}To ensure consistency with CodeScore\textsubscript{test}(n=854), we build a balanced LoCaL dataset, LoCaL\textsubscript{balanced} of 854 entries (427 mutation and 427 optimization variants), and evaluate its SFD and DFS region coverage against CodeScore\textsubscript{test}. We run Algorithm \ref{alg:mse_gap} on the union of LoCaL\textsubscript{balanced} and CodeScore\textsubscript{test} which yields $x_{lo}=0.65, x_{hi}=0.90, y_{lo}=0.10 \text{ and }y_{hi}=0.90$ as the threshold values for SFD and DFS regions. We then plot SurfaceSim (Eq. \ref{eq:surface_similarity}) vs $df_{score}$ for both LoCaL\textsubscript{balanced} and CodeScore\textsubscript{test} in Fig. \ref{fig:RQ3_results}. 
\subsubsection{Results and Analysis} As illustrated in Fig. \ref{fig:RQ3_results}, LoCaL\textsubscript{balanced} is densely populated around the corner regions (SFD and DFS), whereas $CodeScore_{test}$ clusters around the Control region. LoCaL places \textbf{216 samples in DFS (25.3\%)} and \textbf{249 in SFD (29.2\%)}. On the other hand, CodeScore\textsubscript{test} has only \textbf{59 samples in DFS (6.9\%)} and \textbf{none (0\%)} in SFD, providing limited coverage of challenging corner regions for CEMs. Even when LoCaL points don't fall strictly inside the DFS or SFD region, they tend to concentrate around the corners. 

On the other hand, CodeScore's distribution is dominated by pairs with low surface similarity and low $df_{score}$. 
We substantiate this by computing, for each dataset, the average distance from every control-region point to its nearest SFD/DFS boundary. LoCaL’s control points average 0.12 away from an SFD or DFS boundary, whereas the baseline averages 0.24, roughly twice as far. This shows LoCaL points stay much closer to the difficult corner regions, even when they don’t cross the boundary line.
The empty white space to the left (Fig. \ref{fig:RQ3_results}) indicates that very few pairs in either dataset have a surface similarity lower than 0.4. This is expected since the pairs here are different implementations of the same problem written in the same language, which naturally results in a minimum level of surface overlap.  
\par Although the thresholds derived in this section depend on the data used to fit them and can shift under different sampling, they provide decision boundaries that allow analysis of the hit rate of LoCaL's mutated and optimized variants in the SFD and DFS regions, respectively. Experiments demonstrate that out of the 2652 mutated variants in LoCaL, \textbf{1586 (59.80\%)} fall within the SFD region (Fig. \ref{fig:RQ3_results}). For the optimized variants, \textbf{237 (50.97\%)} out of 465 fall within the DFS region. 
\begin{rqbox}
\textbf{Findings of RQ3:} Prior work fails to cover the error-prone regions for CEMs, with only 6.9\% of samples falling in the DFS region and 0\% in the SFD region. LoCaL substantially increases coverage by placing \textbf{54.5\%} of the total samples strictly inside the DFS and SFD regions.
\end{rqbox}

\subsection{RQ4: Assessing the Effectivenss of LoCaL}
\label{sec:RQ4}
\subsubsection{Experimental Setup} This research question investigates how well LoCaL can identify the limitations of the CEMs considered. Because there is no other available Python benchmark besides CodeScore\textsubscript{test} to compare against LoCaL, we also evaluate the CEMs on the Java benchmark, ShareCode\textsubscript{java}. However, it comprises functionally equivalent and non-equivalent solution pairs drawn from \textit{ShareCode.io} \cite{sharecode}, with only binary (0/1) labels rather than continuous scores. It also cannot be used to evaluate CodeScore, the Python-only CEM. Moreover, ShareCode\textsubscript{java} cannot be directly compared to LoCaL since CEM behavior can be language-dependent, and Java codes inherently have more syntactic overlap than Python.  
\subsubsection{Results and Analysis} The results are presented in Table \ref{tab:RQ4_results}. All four CEMs perform significantly worse on LoCaL compared to CodeScore\textsubscript{test}. The largest increase in prediction error is shown by CodeScore \textbf{(9.38 times)}, followed by CodeBLEU \textbf{(2.96 times)}, CrystalBLEU \textbf{(2.70 times)}, and CodeBERTScore \textbf{(1.39 times)}. We also report the performance of the CEMs on the mutation-only and optimization-only subsets of LoCaL, isolating the impact of each variant type. As shown in Table \ref{tab:RQ4_results}, each CEM, except CodeBERTScore, shows performance degradation on both subsets of LoCaL.  
CodeBERTScore performs better on the optimization-only subset because it always gives a high score (an average of 0.91 on LoCaL) regardless of functional similarity (also discussed in Section \ref{RQ2}). Since optimization variants have ground-truth functional similarity close to 1, CodeBERTScore ``appears" to be effective on the optimization-only subset. However, its extremely poor performance on the mutation-only subset proves this apparent effectiveness stems from CodeBERTScore's tendency to provide inflated scores rather than true semantic sensitivity. Even against ShareCode\textsubscript{java}, a Java benchmark with greater surface overlap, LoCaL remains the tougher benchmark (Table \ref{tab:RQ4_results}), which further highlights its capability to challenge CEMs. This gain by LoCaL can be attributed to our two-phase strategy: first isolating the error-prone regions and then curating pairs that target those regions. Moreover, the lack of an effective benchmark in Java motivates a Java version of LoCaL, enabling stronger evaluation of CEMs in multiple languages. 
\par From the LoCaL samples, we pick 100 cases with the largest errors for each metric. We then analyze these cases to identify error-inducing patterns. The clearest pattern, for all metrics, is the large gap between surface similarity and ground truth functional similarity, with an average absolute difference of 0.86. This noticeable and consistent divergence between surface-level features and actual functionality in the most erroneous samples reinforces our findings in RQ1 and RQ2. We also observe that all four metrics perform worst on the samples drawn from APPS, followed by PIE, both of which consist of program-level code pairs. This indicates reference-based CEMs struggle more with program-level pairs compared to method-level pairs. 
\begin{table}[t]
\centering
\caption{Mean Absolute Error (MAE, Eq. \ref{eq:mae_k}) for four CEMs across five datasets. Here, LoCaL\textsubscript{mut} and LoCaL \textsubscript{opt} indicate the mutation-only and the optimization-only subsets of LoCaL, respectively.  }
\vspace{-3mm}
\label{tab:RQ4_results}
\renewcommand{\arraystretch}{1.0}
\small
\setlength{\tabcolsep}{5pt}
\begin{tabular}{l|l|c|c|c|c}
\hline
\textbf{Language} &
{\diagbox[width=6.8em]{\textbf{Dataset}}{\raisebox{-1.0ex}{\textbf{CEM}}}} &
\textbf{CodeBLEU} &
\textbf{CrystalBLEU} &
\textbf{CodeBERTScore} &
\textbf{CodeScore} \\
\cmidrule(lr){1-6}
\multirow{4}{*}{Python}
  & CodeScore\textsubscript{test} & 0.27 & 0.27 & 0.57 & 0.08 \\
  \cmidrule(lr){2-6}
  & \textbf{LoCaL}                & \textbf{0.80} & \textbf{0.73} & \textbf{0.79} & \textbf{0.75} \\
\cmidrule(lr){2-6}
  & LoCaL\textsubscript{opt}      & 0.67 & 0.87 & 0.16 & 0.36 \\
\cmidrule(lr){2-6}
  & LoCaL\textsubscript{mut}      & 0.83 & 0.71 & 0.90 & 0.83 \\

\cmidrule(lr){1-6}
\multirow{1}{*}{Java}
  & ShareCode\textsubscript{java} & 0.43 & 0.47 & 0.46 & N/A \\
\hline
\end{tabular}
\end{table}

\begin{rqbox}
\textbf{Findings of RQ4:} LoCaL is highly effective in revealing the weakness of existing reference-based static CEMs. The most challenging LoCaL samples for all the CEMs consistently show a large gap between functional and surface similarity, further supporting the RQ1 and RQ2 findings that \textit{surface bias} inhibits these metrics.
\end{rqbox}
\subsection{RQ5: Impact of Incorporating LoCaL into Training}
\label{sec:RQ5}
As seen in Table \ref{tab:RQ4_results}, CodeScore's performance on its inference dataset is an impressive $0.08$ MAE, but it does not retain its performance on the unseen data from LoCaL. This motivated us to ask the question: \textit{Can LoCaL samples incorporated in training help CodeScore perform better on the error-prone regions?} 
To answer, we perform an exploratory study by systematically varying the percentage of LoCaL samples in the training set of CodeScore and analyzing its performance on unseen data. 
LoCaL is designed as an evaluation dataset for CEMs, not a training dataset, so our study investigates smaller scale CodeScore training that is able to 
provide an initial answer to the RQ, and we leave a full study of diversity
in CEM training corpora to future work.

\subsubsection{Experimental Setup} First, we randomly pick 2500 entries from the original training set of CodeScore and 312 entries from the validation set, preserving the original 12.5 \% split. We keep the scale of this study small since integrating  LoCaL samples in the training set requires converting the code pairs used in the original training to our $df_{score}$ so that all data share the same ground-truth label definition.  This conversion needs a substantial amount of time as each pair is evaluated on 1,000 test inputs with verified constraints. After the conversion (with timeouts removed), we get 2026 entries in the training set and 259 entries in the validation set. Training CodeScore on this data yields the model we call CodeScore\textsubscript{base}. We then keep the training set size fixed at 2026 and vary the percentage of LoCaL, $p$, from 0\% to 100\% in 10\% increments by replacing original training samples with LoCaL. For each setting, we retrain CodeScore on this mixed set and denote the retrained models as CodeScore\textsubscript{mixed}. The PIE samples in LoCaL are held out for inference, and the remaining LoCaL samples are randomly included in training.
\par We evaluate each model on three sets: two disjoint sets of 259 (same as the validation set size) random samples each: (i) Baseline (from the original inference set of CodeScore), (ii) LoCaL\textsubscript{test} (from the held-out PIE samples of LoCaL), and (iii) their combined set. For each $p$, we train five independent models and report the $\text{mean}(\pm\text{standard deviation})$ of MAE (Table \ref{tab:RQ5_results}). The table includes MAE on the combined set and on each set separately, letting us isolate the impact of incorporating LoCaL on both the baseline and LoCaL\textsubscript{test}.
Multiple runs help control the randomness arising from model initialization, data sampling, etc., and provide confidence that the trends observed are not artifacts of a single ``lucky" run. Model hyperparameters are held constant across $p$. 
\subsubsection{Results and Analysis} 

Table \ref{tab:RQ5_results} shows a consistent drop in the combined MAE as LoCaL percentage in training increases, with the LoCaL-only model $(p=100\%)$ showing a drop of $\approx40\%$ , relative to the no-LoCaL$(p=0\%)$ model. Importantly, even a small percentage of LoCaL samples $(10-20\%)$ in training results in a large improvement for CodeScore. Breaking down to individual datasets, we see LoCaL\textsubscript{test} MAE sharply drops, then plateaus. On the other hand, interestingly enough, baseline MAE seems largely invariant to $p$ and almost remains stable. We can see the pattern at the endpoints. When there is no LoCaL data ($p=0\%$) in training, the base model performs poorly on  LoCaL\textsubscript{test}. On the other hand, the LoCaL-only model $(p=100\%)$ performs significantly better on  LoCaL\textsubscript{test} and remains reasonably competitive on the baseline. Overall, \textbf{LoCaL-only training reduces the LoCaL\textsubscript{test} MAE by $\mathbf{\approx85\%}$ with just $\mathbf{\approx 9.1\%}$ increase in the baseline MAE}. This shows that the improvement on the combined dataset comes from the improvement on LoCaL\textsubscript{test}, and also highlights LoCaL's utility as training data. Moreover, the relatively minor performance drop on the baseline as $p$ increases invites further investigation. If the apparent stability on the baseline is largely due to the pretrained knowledge of the underlying model, then fine-tuning only or mostly on LoCaL-like data could be a practical path to robust CEMs. Testing the limits of this claim is an important direction for future work.
\par Although there is some variance at some points in Table \ref{tab:RQ5_results}, the downward trend on LoCaL\textsubscript{test} with higher $p$ is pretty clear. We suspect this variance comes from the lack of data and plan to consider an extended LoCaL-like dataset to solidify the findings of this research question.
\begin{table}[t]
\centering
\caption{Effect of varying LoCaL percentage, $p$ on the MAE of CEMs. Here, $p=0$ denotes CodeScore\textsubscript{base}, and $p>0$ denotes the CodeScore\textsubscript{mixed} model that has $p\%$ LoCaL samples}
\vspace{-3mm}
\label{tab:RQ5_results}
\renewcommand{\arraystretch}{1}
\setlength{\tabcolsep}{3.5pt}
\small
\resizebox{\textwidth}{!}{
\begin{tabular}{l|c|c|c|c|c|c|c|c|c|c|c}
\toprule
\multicolumn{1}{l|}{\diagbox[width=6.8em]{\textbf{Dataset}}{\raisebox{-1.0ex}{$\mathbf{p}$ (\textbf{\%})}}}
  & 0 & 10 & 20 & 30 & 40 & 50 & 60 & 70 & 80 & 90 & 100 \\
\midrule

\textbf{Combined}
  & 0.45 & 0.31 & 0.28 & 0.26 & 0.27 & 0.26 & 0.26 & 0.26 & 0.26 & 0.26 & 0.27 \\
\scriptsize ($\pm$Std.)
  & \scriptsize $(\pm0.04)$ & \scriptsize $(\pm0.02)$ & \scriptsize $(\pm0.01)$ & \scriptsize $(\pm0.01)$ & \scriptsize  $(\pm0.02)$ & \scriptsize $(\pm0.01)$ & \scriptsize $(\pm0.01)$ & \scriptsize $(\pm0.01)$ & \scriptsize $(\pm0.02)$ & \scriptsize $(\pm0.02)$ & \scriptsize $(\pm0.01)$ \\
\midrule

\textbf{Baseline}
  & 0.44 & 0.41 & 0.42 & 0.40 & 0.42 & 0.40 & 0.42 & 0.42 & 0.41 & 0.43 & 0.48 \\
\scriptsize (Std.)
  & \scriptsize $(\pm0.03)$ & \scriptsize $(\pm0.01)$ & \scriptsize $(\pm0.01)$ & \scriptsize $(\pm0.01)$ & \scriptsize $(\pm0.01)$ & \scriptsize $(\pm0.01)$ & \scriptsize $(\pm0.01)$& \scriptsize $(\pm0.02)$& \scriptsize $(\pm0.02)$ & \scriptsize $(\pm0.02)$ & \scriptsize $(\pm0.02)$ \\

\midrule

\textbf{LoCaL\textsubscript{test}}
  & 0.46 & 0.22 & 0.15 & 0.12 & 0.13 & 0.11 & 0.09 & 0.09 & 0.10 & 0.09 & 0.07 \\
\scriptsize (Std.)
  & \scriptsize $(\pm0.05)$ & \scriptsize $(\pm0.05)$ & \scriptsize $(\pm0.02)$ & \scriptsize $(\pm0.02)$ & \scriptsize $(\pm0.02)$ & \scriptsize $(\pm0.01)$ & \scriptsize $(\pm0.02)$ & \scriptsize $(\pm0.01)$ & \scriptsize $(\pm0.03)$ & \scriptsize $(\pm0.02)$ & \scriptsize $(\pm0.00)$ \\

\bottomrule
\end{tabular}

}
\end{table}

\begin{rqbox}
\textbf{Findings of RQ5:} When LoCaL samples are mixed into training (even in small proportions), CodeScore performs remarkably better on held-out LoCaL data while retaining almost the same performance on its original test split. 

\end{rqbox}
\section{Threats to Validity}
Our findings may be affected by both internal and external threats.

\textbf{Threats to Internal Validity} In our study, we use different tools and scripts to automate data analysis and the multi-step process of constructing LoCaL (Fig. \ref{fig:local-pipeline}). While we cannot rule out implementation bugs, we try to mitigate the risk by using well-established tools like  MutPy and Atheris and by repeating each experiment multiple times. Since there is always a chance of not capturing the best performance of CEMs or baselines, we mimic the original setting of each CEM as much as possible. Moreover, we compare LoCaL with the baseline only after ensuring the same ground-truth definition. The scores in LoCaL are derived from fuzzing, which introduces inherent randomness. To address this, we run each fuzzing process five times and also manually verify the input constraints to preserve a defined input space. This gives us confidence in the reliability of LoCaL's functional similarity scores. We evaluate CEMs on code pairs with sharp contrast between functionality and appearance. To address concerns about real-world relevance, we use optimizations and mutations, well-established practices in software engineering, rather than arbitrary code edits that could generate more challenging but impractical pairs. While we find that incorporating LoCaL samples into training can improve CEM performance, we acknowledge that this finding might be threatened by the small size of the dataset used.

\textbf{Threats to External Validity}
The thresholds that we use to define SFD and DFS regions (Section \ref{RQ3}) may vary across datasets and hence should be viewed as dataset-calibrated partitions rather than universal ground truth. Despite this, Algorithm \ref{alg:mse_gap} can provide a reasonable estimation of the error-prone corner regions in any dataset. While we select the four Code Evaluation Metrics (CEMs) based on a literature survey of usage trends, there might be newer CEMs that perform better on LoCaL. Since LoCaL currently consists of Python pairs only, our findings of the research questions may not generalize directly to other programming languages. Nonetheless, the methodology behind LoCaL can be adapted to build benchmarks in different programming languages. 

\section{Conclusion}
In this work, we quantify the impact of \textit{surface bias} on reference-based static code evaluation metrics (CEMs), demonstrating that these metrics mostly struggle with code pairs that have a significant gap between their surface similarity and functional similarity. Such scenarios, however, have not been considered in prior CEM development or evaluation. To address the gap, we propose \textbf{LoCaL}, a large-scale Python benchmark that provides continuous functional similarity scores to code pairs. LoCaL builds on a novel differential fuzzing-based strategy that dynamically generates thousands of test cases and executes them to compute reliable similarity scores. We find that LoCaL effectively hits the error-prone regions and causes significant performance degradation in the CEMs analyzed. Our exploratory study also suggests incorporating LoCaL samples into training can enhance CEM's performance, reducing its bias towards surface-level features. Beyond serving as an evaluation framework for CEMs, LoCaL provides reusable ground-truth similarity scores for downstream tasks like code optimization, code clone detection, code refactoring, and automated bug repair. 
\par As the first benchmark explicitly designed to counter \textit{surface bias}, LoCaL paves the way for code evaluation metrics that do not merely reward surface resemblance but instead reflect true program behavior. Looking forward, we hope LoCaL will serve as both an evaluation benchmark and a reminder that LoCaL-like data must be considered in the design of code evaluation metrics that seek to assess functional correctness without execution.

\bibliographystyle{ACM-Reference-Format}
\bibliography{ref}

\end{document}